# Towards Providing Low-Risk and Economically Feasible Network Data Transfer Services

MUGUREL IONUT ANDREICA
Computer Science and Engineering Department
Politehnica University of Bucharest
Splaiul Independentei 313, sector 6, Bucharest
ROMANIA
mugurel.andreica@cs.pub.ro    https://mail.cs.pub.ro/~mugurel.andreica

VASILE DEAC, STELIAN TIPA
Faculty of Management
The Bucharest Academy of Economic Studies
Piata Romana 6, sector 1, Bucharest
ROMANIA
deac_vasile@yahoo.com, stelian.tipa@intesasanpaolo.ro

*Abstract:* - In the first part of this paper we present the first steps towards providing low-risk and economically feasible network data transfer services. We introduce three types of data transfer services and present general guidelines and algorithms for managing service prices, risks and schedules. In the second part of the paper we solve two packet scheduling cost optimization problems and present efficient algorithms for identifying maximum weight (k-level-) caterpillar subtrees in tree networks.

## 1 Introduction

The need for efficient network data transfer services has increased rapidly during the past few years and it continues to have an increasing trend. This is in correlation with the increases in the world-wide volumes of data produced per year and in the overall Internet bandwidth and number of users. Many business domains make use of large data volumes which need to be accessed, processed and transferred. Recent video on demand and live streaming applications deliver multimedia streams of variable quality to consumers, thus requiring large transfer speeds (bandwidth). Banks and other financial institutions need to store and process large amounts of data, generated by the transactions of their customers. The transfer of the data must be performed reliably (no errors can be accepted) and quickly (in order to perform real-time data processing). Large distributed systems (e.g. Grids), run applications which consist of workflows, in which the output generated by a part of the application must be fed as input to another part of the application. Thus, data transfer scheduling techniques are required in order to obtain increased performance. In this paper we take the first steps towards providing low-risk and economically feasible network data transfer services. In Section 2 we discuss the types of services that could be provided by a data transfer service provider. In Section 3 we consider scheduling details, pricing policies and risk mitigation methods, such that the data transfer services present low economic risk and financial feasibility. In Section 4 we discuss two packet scheduling optimization problems. In Section 5 we present algorithms for identifying maximum weight (k-level-) caterpillars in trees. In Section 6 we discuss related work and we conclude.

## 2 Types of Data Transfer Services

The context in which dedicated data transfer services can be provided is that in which the entire network infrastructure is owned by the service provider. Thus, we define the network as a graph composed of *n* vertices and *m* edges. A vertex is either a user computer or the device through which the user connects to the network infrastructure of the provider (e.g. a cable modem or ADSL modem), a router, a switch, or any other network device. An edge between two vertices corresponds to a physical link between the two vertices and has an associated latency and bandwidth (if we have a full-duplex link, then we need a latency and bandwidth value for each direction). We only consider point-to-point links and we ignore shared network media (e.g. Ethernet on a bus topology, or wireless) on purpose. Note that the provider does not need to map its entire network in this graph. In this case, a vertex may stand for a group of network devices or for part of a network. However, the physical characteristics of the network links (graph edges) should be adjusted accordingly (e.g. an edge between two vertices *i* and *j* may now stand for multiple physical links between the network devices of the groups corresponding to the two vertices; in such a case, the bandwidth for each direction of the edge could be the sum of the bandwidths of the physical network links for that direction, but it's difficult to say what the "combined" latency should be). We consider that all the traffic that exists in the network is the result of using some of the data transfer services which we describe next. The most common type of data transfer service, which is provided by most Internet Service Providers (ISPs) nowadays, is the following. They guarantee a maximum upload and download bandwidth to the users and make no guarantees for anything else. The user can choose a quality level – the larger the quality level, the larger the upload and download bandwidths are. The users are charged a flat fee $F(j)$ (for quality level $j$) per month as long as a combination $C$ of the total upload traffic $U$ and download traffic $D$ (during the current month) does not exceed an upper bound $B$ ($C$ may be $U+D$, or some other function chosen by the provider). Then, the users are charged an extra fee, usually directly proportional with the excess: $E(j) \cdot (C-B)$. This business model works well, but it is not useful in any of the situations presented in Section 1. This is because these are only best-effort services and provide no end-to-end guarantees. The following types of services focus on providing end-to-end guarantees. The best model for using such services is that in which the users (or user applications) submit data transfer requests to a central scheduler (like in [2, 3]). The first type of requests is given by *fixed bandwidth-fixed duration* requests. Such a request has the following parameters: $(t_1, t_2, B, D, lmax, s, f)$, meaning that the data transfer requires a minimum amount of bandwidth $B$ for a duration of time $D$, the earliest possible starting time of the transfer is $t_1$, the latest possible finish time is $t_2$, and the transfer takes place between nodes $s$ and $f$. The scheduler must assign to the request a (directed) path from $s$ to $f$ in the graph and a time interval $[ts, ts+D]$ in which the required amount of bandwidth is available on each edge on the path (in the direction from the source to the destination) and the sum of latencies of the edges on the path is at most *lmax*. Then, data is transferred on the assigned path and during the assigned time interval at (at most) the requested bandwidth. A second type of requests is given by *fixed data-fixed duration* requests. The parameters of such a request are: $(t_1, t_2, TD, dataid, s, f, o)$, meaning that it needs to transfer *TD* bytes of data, the transfer takes place between nodes *s* and *f*, and the transfer must occur between time moments $t_1$ and $t_2$, but the transfer speed does not have to be constant (i.e. it can vary in time; all that matters is that the moment *tf* when the last byte of data arrives at the destination is $\leq t_2$, and we do not start transferring the first byte of data before $t_1$). The data is initially located at the source node *s* and is identified by an identifier *dataid* (e.g. file name, location and offset within the file). The data does not have to be transferred on a single path – this is up to the scheduler; but the transfer must be reliable (i.e. all the data must reach its destination). The *o* parameter is a boolean flag which indicates if the data must be received in order (*o=true*) or if it can be reordered at the destination (*o=false*) by adding extra information regarding its position in the flow to every transferred packet. Another type of requests is given by *fixed bandwidth-variable duration* requests. The parameters of such a request are: $(t_1, B, lmax, s, f)$, meaning that the scheduler needs to allocate to the request a (directed) path from *s* to *f*, starting at time moment $t_1$ and for an indefinite duration. A bandwidth of at least *B* must be available on every edge of the path (in the corresponding direction, from *s* to *f*) and the sum of the latencies of the edges on the path must be at most *lmax*. So far, we only considered independent and unrelated requests. However, the two types of *fixed duration* requests can be extended as follows. The scheduler receives a group of *M* requests $(r(1), ..., r(M))$ which are connected by precedence constraints. To be more precise, besides satisfying the constraints of each request, we are also given a directed acyclic graph (DAG) which has a vertex *i* for every request *r(i)*. We have a directed edge from vertex *i* to vertex *j* if the data transfer corresponding to request *r(j)* must start only after the data transfer corresponding to request *r(i)* is complete. The group may contain both *fixed bandwidth-* and *fixed data- fixed duration* requests.

## 3 Managing Schedules, Prices and Risks

Economic feasibility is strongly related to the profits the data transfer service provider can obtain from its customers. Thus, the pricing policy plays an important role. Depending on the (expected) number of customers, the provider may choose a fixed pricing policy or may negotiate the price for every data transfer request (it may even present multiple alternatives with different prices to the customer). In the case of *fixed bandwidth-fixed duration* requests, the price should be proportional to *B* and *D*, and inversely proportional to the *slack* $(t_2-t_1-D)$ and *lmax*. That is, the larger the requested bandwidth and duration are, the larger the price should be, and the smaller the upper bound on latency and the slack are, the larger the price should be. Prices may also depend on the time interval $[t_1,t_2]$. If it is difficult to find a time interval of length *D* where a path satisfying the Quality-of-Service (QoS) constraints exists, then the price should be higher. Moreover, the price can be proportional to the amount of already reserved bandwidth on the edges on the chosen path (the higher the bandwidth of the path is utilized, the higher the price). *Fixed data-fixed duration* requests should be charged proportional to the total transferred data (*TD*), and inversely proportional to the length of the time interval $[t_1,t_2]$. Moreover, if the data should be delivered in an ordered fashion, the price should be higher (as there are more constraints imposed on the provider). *Fixed bandwidth-variable duration* requests are slightly more complicated. Obviously, the price should be proportional to *B*, to the actual duration of using the service, and inversely proportional to *lmax*. A combination of fixed and variable costs could be used here. For instance, if the service is used for a duration of at most *D*, then the price could be *CF*; otherwise, if the usage duration is *D'>D* the price will be *CF+CV·(D'-D)* (*CF* and *CV* depend on the other parameters of the request). Note that not all of the requests may be satisfied, as the provider may not have sufficient resources to accommodate all the requests. When receiving a request, if it can be satisfied, the provider should choose the price also based on the risk that this request may force the rejection of future requests which might bring larger revenues (we consider that once accepted, a request cannot be cancelled or rejected later). *Fixed bandwidth-variable duration* requests present the highest risk, as resources might need to be reserved for a long time in order to make sure that the request is satisfied (if, however, the provider is over-provisioned compared to the actual customer demand, these requests may be the most desirable, as they might use the network resources for larger time intervals and, thus, they may be favoured in some sense). Fixed bandwidth-fixed duration requests present the second highest risk and fixed data-fixed duration requests present the lowest risk (those with unordered data delivery are less risky than those with guaranteed ordered data delivery). However, handling a DAG of a group of requests presents significantly higher risks than handling independent requests. Thus, the provider should use a good risk model, as this will influence its pricing policy. A forecast and a simulation component should be included in the risk model. The forecast component should identify patterns of the parameters of the requests received so far and patterns of behaviour for the fixed bandwidth-variable duration requests (e.g. estimations of the actual durations). The forecast component should be used as follows. Given all the available information regarding the requests and a time interval $[t_1,t_2]$, the forecast component should be able to generate a list of *fake* requests, which it estimates that might be received during the interval $[t_1,t_2]$. Then, when deciding the price of a newly received request, we use the simulation component to estimate the overall revenue if the request were accepted (ignoring its price) and the overall revenue if the request were rejected. The simulation is run for a carefully chosen duration (e.g. it simulates *T* seconds or minutes in the future) and uses as input the list of fake requests estimated by the forecast component for the interval *[present time moment, present time moment+T]*, and the currently scheduled requests. The price of the request should be chosen such that the revenue in case of not accepting the request is not larger than the revenue in the case of accepting request (but ignoring its price) plus the price of the request. The forecast component may also be used differently. It could generate $K≥1$ lists of fake requests for a given time interval $[t_1,t_2]$ and it could assign a probability of occurrence *prob(i)* to every list *i* ($1≤i≤K$). Then, the simulation is run for each of the *K* lists, a revenue *R(K)* is computed for every list and then an expected revenue *ER*=the sum of the values *prob(i)·R(i)* is computed and used (we assume that the sum of the values *prob(i)*, $1≤i≤K$, is *1*). Another important component is the module which performs the scheduling of the data transfer requests. The algorithms used are very important and may have a strong influence on the revenues of the provider, depending on how the requests are scheduled. We will present an algorithm for the case when only *fixed bandwidth-fixed duration* requests are considered. The algorithm is based on events (as opposed to time-slot based algorithms). For each event, the time moment and the value by which the bandwidth is modified (a positive or negative value) are stored, i.e. a pair *(t,dB)*. For each network link (and direction), a list of events is maintained. The time moment of an event in the event list of a network link is represented by the start or finish time of a data transfer. The algorithm starts by generating several candidate paths from *s* to *f*, which satisfy the sum of latencies constraints. Then, for each path, we verify if we can schedule the data transfer request on that path in order to satisfy all the other constraints. We consider a family of greedy algorithms for this case: *First-Fit*, *Last-Fit*, *Best-Fit* and *Worst-Fit*. All the events on all the (directed) network links along the candidate path are initially sorted. Also, for each (original) event *(t,dB)* on a link *l*, an event *(t+D,0)* on the same link is added, where *D* is the duration requested for the data transfer. These events are sorted together with the other events. Then, the sorted events are traversed. During the traversal we maintain for each network link *l* a deque *DQ(l)* and the current available bandwidth *cb(l)* of the link (*cb(l)* is initially the total bandwidth of the link). The deque maintains sorted *(tm=time moment, ab=available bandwidth)* pairs, similar to the deque presented for the Time Slot Groups data structure in [7]. Let's assume that we reached an event *(t,dB)* on a link *l*. First, we consider every link *l'* and: *(1)* while *DQ(l')* contains at least *1* pair and the last pair *lp* from *DQ(l')* has *lp.ab≥cb(l')*, we remove *lp* from

$DQ(l')$ ; (2) we insert the pair $(t,cb(l'))$ at the end of $DQ(l')$ ; (3) while the first pair $fp$ of $DQ(l')$ has $fp.tm \leq t-D$, we remove $fp$ from $DQ(l')$. Then, we increment $cb(l)$ by $dB$. After all these, the first pair $fp$ of $DQ(l')$ of each link $l'$ contains the minimum available bandwidth $AB(l')=fp.ab$ of that link on the interval $[t-D,t]$. If $AB(l')$ is greater than or equal to the required bandwidth $B$ for every network link $l'$ on the candidate path and the interval $[t-D,t]$ is included in $[t_1,t_2]$ (given by the request), then a match is found and we perform the following actions. In the case of the First-Fit algorithm, we return the interval $[t-D,t]$ as the solution (the request can be scheduled on the path). For the Last-Fit Algorithm we just store $t$ into $tfin$. For the Best-Fit and Worst-Fit algorithms we compute $MAB=\min\{AB(l')|l'$ is a link on the candidate path$\}$. These two algorithms will maintain a bandwidth value $MB$, initially equal to $+\infty$ for Best-Fit and $-\infty$ for Worst-Fit. If $MAB<MB$ for Best-Fit ($MAB>MB$ for Worst-Fit) we set $MB=MAB$ and $tfin=t$ (note that $MAB \geq B$, because otherwise we wouldn't have performed these actions). Then, at the end, if anything was stored in $tfin$ (in the case of the Last-Fit, Best-Fit and Worst-Fit algorithms), we return the interval $[tfin-D,tfin]$; otherwise, no solution is found. If we schedule a request on a path between time moments $ts$ and $ts+D$, we insert the events $(ts,-B)$ and $(ts+D,+B)$ in the event list of every link $l$ on the path (for the correct direction). The sorting stage of the events can be performed by maintaining a balanced tree, which contains the next event to occur for each network link. At each step, the event occurring at the earliest time is extracted from the balanced tree (and replaced by the next event on its link). The algorithm stops as soon as the latest finish time of the request is passed by. The overall time complexity is $O(E \cdot (\log(NL)+NL))$, where $E$=the total number of events and $NL$=the total number of links on the candidate path.

## 4 Efficient Packet Scheduling

We consider a sequence of $N$ packets which need to be sent in order. The size of packet $i$ is $x(i)$ bytes. We can transfer any sequence of consecutive packets at a time, with the following restrictions. Let's assume that $S$ is the total size of the packets in the sequence. There are $K$ levels for the data transfer. For every level $j$ we have two parameters: $L(j)$ and $C(j)$ ($L(j+1)>L(j)$ and $C(j+1)>C(j)$). Their meaning is the following. If $S \leq L(1)$ then the transfer cost is $C(1)$. If $L(j-1)<S \leq L(j)$ then the cost of the transfer is $C(j)$ ($2 \leq j \leq K$). We cannot transfer sequences of packets whose total size $S$ exceeds $L(K)$ during a single data transfer. We want to transfer all the $N$ packets, in order, such that the total cost of the data transfers is minimized. We will compute $Cmin(i)$=the minimum total cost for sending the first $i$ packets. We have $Cmin(0)=0$. We will first compute the prefix sums $SP(i)=x(1)+\ldots+x(i)$ ($SP(0)=0$ and $SP(1 \leq i \leq N)=SP(i-1)+x(i)$). We will maintain $K$ data structures $SD(j)$ ($1 \leq j \leq K$). We traverse the $N$ packets in order, from $1$ to $N$. When we reach packet $i$, $SD(1)$ will contain the indices of the packets $q<i$ such that $SP(i)-SP(q) \leq L(1)$, and $SD(2 \leq j \leq K)$ will contain the indices of those packets $q<i$ such that $L(j-1)<SP(i)-SP(q) \leq L(j)$. For every data structure $SD(j)$ we will maintain the indices $left(j)$ and $right(j)$, meaning that the packets in the interval $[left(j),right(j)]$ belong to $SD(j)$. $SD(j)$ will allow the addition of a new packet, the removal of an old packet and finding the minimum value $Cmin(z)$ for $left(j) \leq z \leq right(j)$ (by using some functions called $insert$, $remove$, and $getMin$). Initially (at $i=0$) we have $left(*)=0$ and $right(*)=-1$, and all the data structures $SD(*)$ are empty. Then, we will insert the virtual packet $0$ in $SD(1)$ (setting $right(1)=0$). Let's assume that we reached a packet $i$ ($1 \leq i \leq N$). We will first update all the data structures $SD(j)$, in increasing order of $j$ ($1 \leq j \leq K$). While $SP(i)-SP(left(j))>L(j)$ ($1 \leq j \leq K-1$), we remove the packet $left(j)$ from $SD(j)$ and we insert it into $SD(j+1)$ (then we increase $left(j)$ and $right(j+1)$ by $1$). While $SP(i)-SP(left(K))>L(K)$, we remove the packet $left(K)$ from $SD(K)$ (and then we increment $left(K)$ by $1$). Then, we compute $Cmin(i)=\min\{C(j)+SD(j).getMin()|1 \leq j \leq K\}$. If $SD(j)$ (for any $1 \leq j \leq K$) contains no packet, then $SD(j).getMin()$ returns the value $+\infty$. After computing $Cmin(i)$ we insert packet $i$ into $SD(1)$ (increasing $right(1)$ by $1$). We can implement the $SD(*)$ data structures as heaps, in which case insertions and removals have an $O(\log(N))$ time complexity and finding the minimum value takes $O(1)$ time. The time complexity of the algorithm would be, in this case, $O(N \cdot K \cdot \log(N))$. However, the data structures $SD(j)$ can also be implemented as deques. If $SD(j)$ is a deque, then the elements inside it are pairs $(idx, Cmin(idx))$ and are sorted both according to the index $idx$ and the value $Cmin(idx)$. When we insert a packet $idx$ into $SD(j)$, in fact we insert the pair $(idx, Cmin(idx))$ into the data structure. Before doing this, we remove from the end of $SD(j)$ all the pairs $(q,Cmin(q))$ with $Cmin(q) \geq Cmin(idx)$; only after this will we add the pair $(idx,Cmin(idx))$ at the end of the deque $SD(j)$. When we remove a packet $idx$ from $SD(j)$, we check if the first pair from $SD(j)$ is $(idx, Cmin(idx))$; if $yes$, then we remove that pair from the beginning of the deque; otherwise, we perform no changes. In order to find the minimum element from $SD(j)$, we consider the pair $(idx, Cmin(idx))$ from the beginning of $SD(j)$ and return the value $Cmin(idx)$ (or $+\infty$ if $SD(j)$ is empty). The time complexity of all the operations on a deque $SD(j)$ is $O(N)$, leading to an $O(N \cdot K)$ overall time complexity.

A second packet scheduling problem is the following. We want to transfer $N$ equal-sized packets from a source to the same destination using some of the $P$ available disjoint paths (numbered from *1* to *P*). Every path *i* has three parameters: *cf(i)*, *cv(i)*, *pmax(i)*. *cf(i)* is the fixed cost which needs to be paid in order to use the path *i*, and *cv(i)* is the cost which needs to be paid in order to send *1* packet on path *i*. Thus, in order to send $k \geq 1$ packets on path *i*, we need to pay a cost equal to *pcost(i,k)=cf(i)+k·cv(i)*; *pcost(i,0)=0*. Moreover, we cannot send more than *pmax(i)* packets on path *i*. We want to send all the *N* packets to the destination (in any order) by paying the minimum total sum. We will start by pointing out the following fact. Let's assume that we paid the fixed costs for the paths $q_1, \ldots, q_M$ and let's assume that we have $cv(q_1) \leq cv(q_2) \leq \ldots \leq cv(q_M)$. We will send as many packets as possible on the path $q_1$ (because it has the lowest cost per packet), then as many as possible on the path $q_2$, and so on. Thus, if we pay the fixed costs for *M* paths, then *M-1* of them are used fully (we send the maximum possible number of packets on them) and the last one may be used only partially. Based on this observation, we can use the following pseudo-polynomial dynamic programming algorithm. We first sort the paths such that $cv(1) \leq cv(2) \leq \ldots \leq cv(P)$. We will compute *Cmin(i,j)*=the minimum total cost for sending *j* packets, using some of the first *i* paths, and each path is either fully used or not used at all. We have *Cmin(0,0)=0* and *Cmin(0,j>0)=+∞*. For $i \geq 1$ we have *Cmin(i,0≤j<pmax(i))=Cmin(i-1,j)* and *Cmin(i, pmax(i)≤j≤N)=min{Cmin(i-1,j), Cmin(i-1,j-pmax(i))+ pcost(i,pmax(i))}*. During the second stage of the algorithm we traverse the paths from *P* down to *1* and we will compute *Cmin2(i,j)*=the minimum cost of sending *j* packets on a single path, considering only the paths *i, i+1, …, P*. *Cmin2(P+1,0)=0* and *Cmin2(P+1,j≥1)=+∞*. *Cmin2(i,0≤j≤min{N,pmax(i)})= min{Cmin2(i+1,j), pcost(i,j)}*; *Cmin2(i,pmax(i)<j≤N)= Cmin2(i+1,j)*. Then, in order to find the minimum total sum to be paid for sending all the *N* packets, we compute *Smin=min{Cmin(i,N), min{Cmin(i,j) + Cmin2(i+1,N-j) | 0≤i≤P, 0≤j≤N-1}}*. This solution takes *O(P·N+P·log(P))* time and *O(P·N)* space. We can reduce the space complexity to *O(P+N)* as follows. First, we notice that after computing *Cmin(i,\*)* we do not require the values *Cmin(i'<i,\*)*. As soon as we compute all the values *Cmin(i,\*)* we consider all the candidates *Cmin(i,j)+Cmin2(i+1,N-j)* for *Smin*. However, we need to compute *Cmin2(i+1,N-j)* differently. We can consider all the paths *i+1, …, P* and interpret them as half-lines; half-line *i* has the equation *y(x)=cf(i)+x·cv(i)*. Then, we need to compute the lower envelope of these half-lines (or the upper-envelope if we consider *–cf(i)* and *–cv(i)* instead of *cf(i)* and *cv(i)* for every half-line; linear time algorithms for the upper envelope of half-lines exist). After computing the lower envelope, *Cmin2(i+1,N-j)* is equal to the smallest value of any half-line at *x=N-j*; these values can be computed in *O(log(P))* time: we find the x-interval of the lower envelope which contains the value *x=N-j*, then we access the half-line *q* with the smallest value on that interval and then we compute *pcost(q,x)*. This way, the time complexity increases to $O(P \cdot N \cdot log(P)+P^2)$. We can keep the time complexity down to $O(P \cdot N+P^2)$ by traversing the values *x=N-j* in increasing (or decreasing) order. This way, the x-interval of the next value of *x* is either the interval of the current value of *x*, or the next interval on the lower envelope; thus, we can find the x-interval containing each value *x* in *O(1)* time, as we do not need to use binary search. A simpler *O(P·N+P·log(P))* solution starts by sorting the paths s.t. $cv(1) \geq \ldots \geq cv(P)$. Then, we compute *Cmin'(i,j)* (in increasing order of *i*), such that: *Cmin'(0,0)=0*, *Cmin'(0, 1≤j≤N)=+∞*, *Cmin'(1≤i≤P, 0≤j≤min{pmax(i), N})=min{Cmin'(i-1,j), pcost(i,j)}* and *Cmin'(1≤i≤P, pmax(i)<j≤N)=min{Cmin'(i-1,j), Cmin'(i-1,j-pmax(i)) + pcost(i, pmax(i))}*. The answer is *Cmin'(P,N)*. We can use *O(N)* space, like in the case of the *Cmin(\*,\*)* values.

## 5 Largest Caterpillar in a Tree Network

Tree topologies occur frequently in communication networks, particularly in the case of multicast communication. Some types of trees, which were used in [6] for devising an efficient multicast strategy, are the caterpillar graphs. These graphs consist of a central path of vertices, and each of the vertices on the central path may have any number of leaves as neighbors, plus the (at most) two neighbors on the central path. We will now introduce the more general class of *k-level-caterpillars*. First, we define the concept of *k-level-leaf*. A vertex *i* is a *k-level-leaf* (for $k \geq 1$) if every neighboring vertex *j* is a *k'-level-leaf* for some *k'<k*, except for possibly one neighbor. For *k=1*, a vertex *i* is a *1-level-leaf* if it has degree *1*. No vertex is a *0-level-leaf*. A *k-level-caterpillar* is a graph which consists of a central path of vertices such that every vertex on the central path has only *k'-level-leaves* as neighbors (with *k'≤k*), except for the (at most) *2* neighbors on the central path. Thus, the caterpillar graphs mentioned previously are *1-level-caterpillars* according to this new definition. A *0-level-caterpillar* is simply a path. We first provide two algorithms for deciding if a given graph with *n* vertices is a *k-level-caterpillar*. First, we verify if the given graph is a tree (it must consist of a single connected component and the total number of edges must be *n-1*). Then, the first algorithm works in *k* stages. In the first stage we mark all the leaves *i* with *mark(i)=1* and the other vertices *j* are left unmarked (*mark(j)=+∞*). In the $p^{th}$ stage ($2 \leq p \leq k$), we consider all the unmarked vertices of the tree. Let's consider such a vertex *i* (with *mark(i)=+∞*). We count the number *q* of neighbors *j* of vertex *i* for which *mark(j)≥p*. If $q \leq 1$ then we set *mark(i)=p*. Since every stage takes *O(n)* time, the time complexity of this algorithm is $O(n \cdot k)$. The second algorithm has $O(n)$ time complexity and starts after checking that the given graph is a tree. We compute for every vertex $i$ the value $dmax(i)$=the maximum distance from vertex $i$ to some leaf of the tree. We can use, for instance, the algorithm presented in [5] for computing the center of a tree, which computes all the $dmax(*)$ values in overall $O(n)$ time. Then, for every vertex $i$ we consider all of its neighbors $j$ and set $jmax$=the neighbor $j$ with the largest value $dmax(j)$ among all of vertex $i$'s neighbors. We then set $kmin(i)=1+max\{1+dmax(j)|j$ is a neighbor of $i$ and $j \neq jmax\}$ (we consider $max\{empty\ set\}=0$). Vertex $i$ is a $kmin(i)$-level-leaf. Afterwards, we consider all the vertices $i$ with $kmin(i)>k$ : every such vertex must have at most two neighbors $j$ with $kmin(j)>k$ (its potential 2 neighbors on the central path).

We now consider the following problem: given a tree $T$ with $n$ vertices (numbered from $1$ to $n$), in which every vertex $i$ has a weight $wv(i)$ and every edge $(u,v)$ has a weight $we(u,v)$ (the weights may even be negative), we want to find a $k$-level caterpillar subtree of $T$ whose sum of edge and vertex weights is maximum. We root the tree at any vertex $r$. Thus, $r$ will be the tree's root and it will introduce parent-son relationships between the tree vertices. We will traverse the tree vertices bottom-up (from the leaves towards the root). For every vertex $i$ of the tree we will compute $O_1(i)$=the maximum total weight of a $k$-level-caterpillar subtree whose central path of vertices starts at vertex $i$ and contains only vertices in vertex $i$'s subtree, and $O_2(i)$=the maximum total weight of a $k$-level-caterpillar subtree fully contained in vertex $i$'s subtree (and vertex $i$ may belong to the k-level-caterpillar or not). Besides these $2$ sets of values, we will compute for every vertex $i$ the values $Wmax(i,j)$=the maximum total weight of the edges and vertices of a subtree in which vertex $i$ is a $j$'-level-leaf ($j' \leq j$), and all the other vertices are $j$''-level-leaves (for various $j'' < j'$) and are fully contained in vertex $i$'s subtree. We have $Wmax(i,0)=-\infty$. For $j \geq 1$ we have $Wmax(i,j)=max\{Wmax(i,j-1),\ wv(i)$ plus the sum of the values $max\{0,\ we(i,s(i,q))+Wmax(s(i,q),\ j-1)\}$ with $1 \leq q \leq ns(i)\}$ ($ns(i)$=the number of sons of vertex $i$; $s(i,q)$=the $q^{th}$ son of vertex $i$). Then, the values $O_1(i)$ and $O_2(i)$ are computed as follows. First, we compute the sum $S(i)$, defined as follows: the sum of the values $(max\{we(i,s(i,j))+Wmax(s(i,j),k),\ 0\})$, for $1 \leq j \leq ns(i)$. Then we have $Omax=max\{O_1(s(i,j))+we(i,s(i,j))-max\{we(i,s(i,j))+Wmax(s(i,j),k),\ 0\}|1 \leq j \leq ns(i)\}$ and $jmax$ is the index $j$ for which we obtain $Omax$. $O_1(i)=max\{wv(i)+S(i),\ wv(i)+S(i)+Omax\}$. Then we compute $Omax_2 =max\{O_1(s(i,j)) + we(i,s(i,j)) - max\{we(i,s(i,j)) + Wmax(s(i,j),k),\ 0\}|\ 1 \leq j \leq ns(i),\ j \neq jmax\}$. We have $O_2(i) = max\{O_1(i),\ O_1(i)+Omax_2,\ max\{O_2(s(i,j))|1 \leq j \leq ns(i)\}\}$. $O_2(r)$ is the maximum total weight of a $k$-level-caterpillar subtree. The time complexity is $O(n \cdot k)$.

# 6 Related Work and Conclusions

Efficient data transfer scheduling models and techniques were discussed and presented in [2, 3, 7]. Time-slot based scheduling data structures and algorithms were introduced and analyzed in [3, 7]. Multicast strategies based on caterpillar topologies were discussed in [6]. An economy-based method for scheduling data-intensive applications in Grids was given in [1]. Some risk assessment methodologies were presented in [4].

In the first part of this paper we presented the first steps towards providing low-risk and economically feasible data transfer services. We introduced three new types of data transfer services and discussed methods of setting prices, alleviating risks and managing data transfer schedules. In the second part of the paper we discussed two packet scheduling optimization problems and presented algorithms for identifying maximum weight (k-level-) caterpillar subtrees in trees.